\documentclass[aps,amssymb,amsmath,amsfonts,twocolumn,pra]{revtex4}
\usepackage{graphicx}
\usepackage{bm,bbm}
\usepackage{amsthm}
\usepackage{bbold}
\usepackage{hyperref}
\usepackage{pst-all}
\usepackage{graphics}
\usepackage{epsfig}
\usepackage{color}
\usepackage{epsf}

\newcommand{\be}{\begin{eqnarray}}
\newcommand{\ee}{\end{eqnarray}}

\begin{document}

\title{Spectral density of generalized Wishart matrices
and  free multiplicative convolution}

\author{Wojciech M{\l}otkowski$^1$, Maciej A. Nowak$^{2}$, Karol A. Penson$^3$, Karol {\.Z}yczkowski$^{2,4}$}

\affiliation{
$^1$Institute of Mathematics, University of Wroc{\l}aw, pl. Grunwaldzki 2/4, PL 50-284, Wroc\l{}aw, Poland. \\
$^2$Marian Smoluchowski Institute of Physics and Mark Kac Complex Systems Research Center,  Jagiellonian University, ul. S. \L{}ojasiewicza 11, PL 30-348 Krak\'ow, Poland\\
 {\normalsize\itshape $^{{3}}$ Sorbonne Universit\'{e}s, Universit{\'e} Paris VI,
  Laboratoire de Physique de la Mati{\`e}re Condens{\'e}e (LPTMC), CNRS
  UMR 7600,  t.13, 5{\`e}me {\'e}t. BC.121, 4~pl. Jussieu, 
F 75252 Paris Cedex 05, France }\\
$^4$Center for Theoretical Physics, Polish Academy of Sciences,
al.\ Lotnik\'ow 32/46  PL 02-668 Warszawa, Poland}

\date{June 24, 2015}

\begin{abstract}
We investigate the level density for several 
ensembles of positive random matrices of a Wishart--like structure,
$W=XX^{\dagger}$, 
where $X$ stands for a nonhermitian random matrix.
In particular, making use of the Cauchy transform, we study 
free multiplicative powers
of the Marchenko-Pastur (MP) distribution, ${\rm MP}^{\boxtimes s}$, 
which for an integer $s$ yield Fuss-Catalan distributions corresponding to a product of $s$ independent
square random matrices, $X=X_1\cdots X_s$. 
New formulae for the level densities are derived 
for $s=3$ and $s=1/3$.
Moreover, the level density corresponding to the generalized Bures distribution,
given by the free convolution of arcsine and MP distributions is obtained.
 We also explain the reason of such a curious  convolution. 
The technique proposed here allows for the derivation of the level  densities
 for several other cases.

\end{abstract}
\maketitle

\section{Introduction}

Ensembles of nonhermitian random matrices are of considerable scientific 
interest \cite{Fo10} in view of their numerous applications 
in several fields of statistical and quantum physics \cite{encRM}.
On the other hand, any ensemble of nonhermitian matrices $X$
allows us to write a positive, hermitian matrix of the {\sl Wishart} form,
\begin{equation}
X \to W=\frac{XX^{\dagger}}{{\rm Tr} XX^{\dagger}} \ .
\label{wishart}
\end{equation}
The normalization implies that the random matrix
satisfies a fixed trace condition, Tr$W=1$, 
so it can be interpreted as a density matrix.

Ensembles of such random density matrices analyzed in \cite{ZS01}
can be obtained by taking a random pure state on a bipartite system
and performing a partial trace over a single subsystem. In such a case
of an isotropic,  structureless ensemble of random pure states,
generated according to the unique, unitarily invariant measure, 
the asymptotic level density of the corresponding quantum states is 
described  by the  Marchenko--Pastur (MP) distribution $P_{1,c}$ \cite{MP67}.
Here  the  parameter $c$ is determined by the ratio of the dimensions of the 
auxiliary and the principal quantum systems.

If the global unitary symmetry of the measure defining 
the ensemble of pure random states is broken,
the partial trace yields {\sl structured} ensembles 
of random density matrices. They can be constructed
combining products of non-hermitian random Ginibre matrices
and sums of random unitary matrices distributed according to 
the Haar measure. Investigation of these ensembles
initiated in Ref.~\cite{ZPNC11} was further developed  by Jarosz~\cite{Ja11,Ja12}.

Random matrices described by the Wishart ensemble 
corresponding to the product of $s$ Ginibre matrices,
$X=G_1G_2 \cdots G_s$, were found to be useful in the  description  of the 
level density of mixed quantum states associated with  a graph~\cite{CNZ10}
and states obtained by projection onto the maximally entangled
states of a multi--partite system \cite{ZPNC11}. 
Hence these distributions describe asymptotic statistics of the 
Schmidt coefficients characterizing entanglement of 
a random pure state \cite{ZS01}.

As the moments of the level density $P_s(x)$ for such ensembles
are known to be  asymptotically described
by the Fuss--Catalan numbers \cite{BBCC11,Ml10},
\begin{eqnarray}
C_s(n)=
\frac{1}{sn+1}\binom{sn+n}{n} , \,\,\,\,\,\,\, s>0, 
%
\label{cata1}
\end{eqnarray}
these distributions are called {\sl Fuss-Catalan}.
These distributions describe singular values
of products of independent Ginibre matrices --
see \cite{AKW13,AIK,KZ14},
but they are also known \cite{AGT10} to describe the asymptotic
distribution of singular values of the $s$--power of a 
single random Ginibre matrix $G^s$.

These distributions may be considered as a generalization 
of the  Marchenko-Pastur distribution for square random matrices, $P_1(x)$,
which corresponds to the case $s=1$. The Fuss--Catalan distributions
can be interpreted  as the free multiplicative convolution product \cite{BBCC11}
of $s$ copies of the MP distribution $P_1(x)$,   
i.e. $P_s(x)=[P_1(x)]^{\boxtimes s}$.
The spectral distribution of $P_s(x)$ for a product of an arbitrary number 
of $s$ random Ginibre matrices was  analyzed by Burda et al \cite{BJLNS10}. 
The spectral distribution for the general case  of random rectangular matrices has been also studied ~\cite{AIK,DC14}. 
This 
distribution was expressed as a solution of a polynomial equation and it was conjectured that 
the finite size effects can be described by a simple multiplicative correction.

An explicit form of $P_2(x)$ was derived in Ref.~\cite{PS01} in the context of the 
construction of generalized coherent states from combinatorial sequences.
An exact form of the Fuss-Catalan distributions
for any integer $s$  was derived in Ref.~\cite{PZ11} in terms of
hypergeometric functions $_sF_{s-1}$.
These results were extended in \cite{MPZ12}, in which 
the Mellin transform was used to derive
analogous distributions for  rational values of the exponent $s=p/q$ 
in terms of special functions.
Free multiplicative powers of the MP distributions
were investigated by Haagerup and M{\"o}ller \cite{HM12},
generalized Fuss--Catalan distributions were studied in Ref.~\cite{LSW11,AKW13}
and  power series expansions were recently obtained in Ref.~\cite{DC14}.

The main aim of this work is to derive a wide class of
new results concerning the level density for generalized Wishart 
ensembles of random matrices. We furthermore want to make a case for the  power of free random calculus  in handling  the problems of  quantum information theory.  Application of free random variables calculus to the area of quantum information, advocated  by the authors  already in 2010~(\cite{KAROLTALKS} and \cite{ZPNC11}),
is currently becoming a standard calculational tool. This is one of the developments that  prompted us to  present some old and some new results from the viewpoint  of the  free random variables formalism.
First,  we explain how the theorems for so-called isotropic random matrices  explain curious combinatorial relations for the class of Bures-like measures. Next,  we show that several related results already known in the literature can be put
on the same footing by using the resolvent method
and  the Voiculescu $S$--transform \cite{VDN92}.
An analytical expression for the level density can be obtained,
provided the corresponding Green's 
 forms a polynomial equation of 
a low order.
For instance, the higher order Fuss--Catalan (FC) distribution $P_3(x)$,
originally expressed by special functions \cite{PZ11}
is shown here to be representable  in terms of elementary functions.


The techniques based on the Cauchy transform are applicable
for ensembles of random matrices related to the free convolution 
of  the Marchenko-Pastur distribution $P_1(x)$,
the Arcsine distribution ($AS$), and their free powers. 
In the case of their free product 
one obtains the Bures distribution \cite{SZ04,OSZ10,FK14},
while  higher values of the exponent $s$ lead to its generalization
referred to as the $s$--Bures distribution. It is worth  mentioning
that these distributions belong to the broader class of  
{\sl Raney distributions} studied in Refs.~\cite{PZ11,MPZ12}.

\medskip

This paper is organized as follows.
In section II we review basic properties of the Cauchy transform
and we inspect  how the level density can be derived from the Green's functions.
In section III we cover the  Haagerup-Larsen theorem for large isotropic random matrices  and we demonstrate how one can apply it for the Bures class of measures. 
Section IV  unravels several spectral densities for various powers (including fractional ones) of the Marchenko-Pastur distribution. As an example,  we discuss the Marchenko-Pastur 
distribution $P_{1,c}$ with an arbitrary 
rectangularity parameter $c$ and the arcsine distributions, 
for which the Green's function is given 
as a solution of a  quadratic equation. Furthermore, 
we discuss the generalized Fuss-Catalan distribution $P_{2,c}$
and the generalized Bures distribution, 
for which the Green's function is given by a Cardano 
solution of a cubic equation. 
The third order generalized Fuss-Catalan distribution $P_{3,c}$
and the $2$-Bures distribution are  studied in another subsection. 
In these cases the Green's function is given by a Ferrari solution of a quartic equation,
which allows us to express the corresponding level density in terms of 
elementary functions. Some technical details of the derivations
are relegated to  Appendix~A.

\section{Cauchy functions and level densities}\label{secII}

To derive the level density corresponding to certain ensembles 
of random matrices, and more generally, to some free convolutions
of the Marchenko-Pastur (MP) distribution, we will use the 
Voiculescu $S$-transform and the Cauchy functions.

\medskip 

Consider a square random matrix $X$ of size $N$ pertaining to the Ginibre ensemble
of non-hermitian random matrices. The Wishart matrix $W=XX^{\dagger}$
is positive, and its level density is asymptotically, $N \to \infty$,
described by the Marchenko--Pastur distributions \cite{MP67}, with the rectangularity
parameter $c$ set to unity,
\begin{equation}
P_1(x) =\frac{1}{2 \pi} \sqrt{\frac{4-x}{x}} , \ \ \ 
x\in [0,4].
\label{MP}
\end{equation} 
The variable $x$ denotes a suitably rescaled eigenvalue $\lambda$ of $W$.
If a random Wishart matrix is normalized 
according to the trace condition Tr$W=1$, 
the rescaled variable reads $x=\lambda N$,
which implies that the mean value $\langle x \rangle$ is set to unity.
Thus the MP distribution describes asymptotically
the level density of random quantum states generated with a 
measure induced by the Hilbert-Schmidt metric \cite{ZPNC11}.

In order to analyze convolutions of the MP distribution, 
it is convenient to use its  Voiculescu $S$--transform \cite{VDN92}
defined as a function of a complex variable $w$,
\begin{equation}
S_{MP}(w)=\frac{1}{1+w} .
\label{SMP}
\end{equation} 
The $S$-transform of the multiplicative free convolution is given by the product of the $S$-transforms.
i
For instance,
 the Fuss--Catalan distribution $P_s$ of an integer order $s$ \cite{BBCC11,PZ11},
which corresponds to a product of $s$ independent non-hermitian 
random matrices, $X=X_1\cdots X_s$,
can be written as a multiplicative free convolution of the
Marchenko--Pastur distribution, $P_s(X)=[P_1(x)]^{\boxtimes s}$.
Hence the corresponding $S$ transform reads
$S_{C_s}(w)= [S_{MP}(w)]^s$.

Assume now we are given an $S$--transform  $S(w)$,  
which corresponds to an unknown probability measure at the real axis. 
To infer this measure and the spectral density $\rho(\lambda)$,
we write the $S$--transform as
 $S(w)=\frac{1+w}{w}\chi(w)$, 
where 
\begin{equation}
\frac{1}{\chi(w)} G\Bigl( \frac{1}{\chi(w)} \Bigr)-1=w .
\label{chi}
\end{equation}

To recover the resolvent,  we put 
\be
\frac{1}{\chi(w)} \ = \ z  ,
\ee
which  allows us to write an implicit
solution for  the Green's function $G(z)$,
known also as the {\sl Cauchy} function in the mathematical literature,
\be
G(z)\equiv \frac{1}{N}\left < {\rm tr} \frac{1}{z {\bf 1}_N-M}\right> =\frac{1+w(z)}{z} .
\label{Gz}
\ee
Here $M$ represents a random matrix from the ensemble investigated.
In other words, for any given $S$--transform $S(w)$
the corresponding Green function $G(z)$ defined on the complex plane
is given as a solution of the following algebraic equation
\begin{equation}
zw(z)\; S\bigl( w(z)\bigr) = 1+w(z) .
\label{wz}  
\end{equation}
Note that the Green's function Eq.~(\ref{Gz}) acts as a generating function for  the spectral moments $m_k=\frac{1}{N}\left< {\rm Tr} M^k\right> =\int d \lambda \lambda^k \rho(\lambda)$, i.e. 
$G(z)=\sum_{k=0}^{\infty} m_k /z^{k+1}$, as seen by expanding the Green's function at $z=\infty$. 
Another useful function is the Voiculescu R-transform, defined as a generating function for the free cumulants $\kappa_k$, i.e. $R(z)=\sum_{k=1}^{\infty} \kappa_k z^{k-1}$. Both functions $G$ and $R$ are  related by  functional relations $R(G(z))+1/G(z)=z$ (or equivalently  $G(R(z)+1/z)=z$). 
Finally, $R$ and $S$ transforms can be  also related, namely the function   $z=yS(y)$ is the composition inverse of  $y=zR(z)$ and vice versa, when expected value is nonzero~\cite{BURJANNOW}.

For  several cases, Eq.~ (\ref{wz})  can be solved 
analytically with respect to $w$. This is, for instance,  the case 
for the Fuss--Catalan distribution, as $S(w)$ reads
$(1+w)^{-s}$  and Eq.(\ref{wz}) yields a polynomial equation of order $s+1$.
It can be solved analytically for $s=2$ and $s=3$.

Thus to obtain the spectral density we apply the Stieltjes inversion formula.
One needs to analyze all solutions of Eq.(\ref{wz})
to extract the desired information. In the case  $s=2$
the corresponding polynomial has three solutions, one of which is real, 
the remaining pair is mutually complex-conjugated.
On the basis of  the Sochocki-Plemelj formula,
  $\frac{1}{\lambda \pm i \epsilon} ={\rm P.V.}
 \frac{1}{\lambda} \mp i \pi \delta({\lambda})$,  the
negative imaginary part of the Green's function  yields 
 the spectral function
\begin{equation}
\rho(\lambda)=-\frac{1}{\pi} \lim_{\epsilon \rightarrow 0} \Im G(z)|_{z=\lambda+i \epsilon} .
\label{rho}
\end{equation}

As analytic solutions of equations of order three and four
contain  square roots  raised to
powers $1/3$ and $-1/3$,  care has to be taken
by evaluation of the imaginary part of a complex
solution along the real axis - for more details see Appendix A.

We would like to mention, that the relevant spectral function can be  recovered as well from the real part of the resolvent. In this case one uses  the maximal entropy argument, yielding  
\be
\lim_{\epsilon \rightarrow 0} [G(\lambda +i\epsilon) + G(\lambda-i\epsilon)]=
\frac{\partial V(\lambda)}{\partial \lambda},
\label{RealG}
\ee
where $V$  is the random matrix potential  defining the measure, i.e. 
$d\mu(M)= dM \exp (- N {\rm tr} V(M))$ -- see Eynard \cite{Ey01}.

 On the basis of the aforementioned Sochocki-Plemelj formula, the resulting
 equation is a singular integro-differential equation. 
 In the case of the spectral  support localized on a single, 
finite interval, one can solve the equation,  e.g. by methods developed by Tricomi~\cite{TRICOMI}. Interestingly, one can also view
Eq.~(\ref{RealG}) as an equation for the potential $V$,
 provided the spectral density $\rho(\lambda)$ is known.
Then the calculation of the Hilbert transform of the spectral density 
according to Eq.~(\ref{RealG}) yields the derivative of the potential, which after integrating the derivative and using  the rotational 
invariance allows one to infer the form of  $V(M)$.
 The  above procedure, although well-defined, is complicated at the
 technical level. In particular, in the case of the spectral 
functions resulting from the solution of cubic or 
 quartic algebraic equations, integration  yields  complicated
 expressions for $V(M)$, which in general are non-polynomial.

\section{Isotropic random matrices and Bures measure}
We define an isotropic random matrix $X$ as an $N$ by $N$ matrix   having a polar decomposition $X=PU$ where $P$ is a positive semi-definite Hermitian random matrix and $U$ is a unitary random matrix distributed with  the Haar measure.   Such matrices  have  a  spectrum independent of the polar angle on the complex plane. In the large $N$ limit, the  powerful Haagerup-Larsen theorem  holds~\cite{HAAGLAR},  which allows one to infer the radial spectral density of the operator $X$ directly from the spectral properties of the operator $P^2$, provided  $P$ and $U$ are mutually free. 
In the mathematical literature, the case of infinitely large matrices possesing the above feature is called R-diagonal~\cite{RDIAG}.  
An important  consequence of the Haagerup-Larsen theorem is the so-called  "single-ring theorem"~\cite{SINGLERING},  stating   that the radially symmetric spectrum of isotropic random operators is always confined to the 
ring, with known and analytically calculable radii. In particular, the inner radius can be equal to zero, therefore the spectrum of isotropic, large random matrices is always either concentrated on the disc or on the ring. Explicitly, the Haagerup-Larsen theorem says
\begin{eqnarray}
S_{P^2}(F_X(r)-1)=\frac{1}{r^2}
\label{hltheorem}
\end{eqnarray}
where $F_X(r)$ is  the cumulative radial density for the complex eigenvalues of $X$, i.e.
\begin{eqnarray}
F_X(r)=2 \pi \int_0^r ds s \rho_X(s)
\label{cumradial}
\end{eqnarray}
and $S_{P^2}(z)$ is a  Voiculescu S-transform for the operator $P^2$ as defined in the previous section.  

The aforementioned  Bures measures and their generalizations belong to the realm  of applicability of the Haagerup-Larsen theorem. 
Recently, a curious relation has been observed in~Ref.~\cite{Ml10}, stating  that the Bures measure is a free multiplication of the arcsine measure and the Marchenko-Pastur measure. 
The relation has been observed by studying the combinatorial properties of the Bures measure, i.e. was inferred from  some a priori unexpected relations between pertinent moments.  
As far as we know, the general proof  of why such factorization holds is  missing. In this section, we provide a simple argument, showing that such a feature is just a consequence of the Haagerup-Larsen theorem.  
First, let us observe that in the large $N$ limit,  we can  
 write down the averages of the products and ratios of  the operators as the corresponding 
products and ratios of the averages of the operators.  This means, that one can perform all calculations
 of the spectral  properties ignoring the normalization ${\rm Tr} XX^{\dagger}$ in the denominator of Eq.~(\ref{wishart}), 
and then, at the end of calculations,  perform the rescaling of the argument of the resolvent by the value
 $a=\frac{1}{N} \langle {\rm tr} XX^{\dagger} \rangle $, according to the relation
\begin{eqnarray}
G_{\frac{P}{a}}(z) = \frac{1}{N}\left< {\rm tr} \frac{1}{z-\frac{P}{a}}\right>  =aG_P(za).
\label{resc}
\end{eqnarray}
The Bures measure is constructed  from the Wishart measure (modulo above-mentioned normalization)   as $X=(U_1+U_2)G$,  where $U_i$ are Haar-distributed  unitary matrices  measures and $G$ is a Ginibre matrix.  Both ingredients   of the above product fulfill in the large $N$ limit the assumptions of the Haagerup-Larsen theorem, i.e  $U_1+U_2=P_U U$, where $P_U=|U_1+U_2|$ and $G=P_G U^{'}$, where $P_G$ is the positive part of the Wigner semicircle (Wigner's semiquarter), and $U$  and $U^{'}$ are Haar measures. 
The square of the $P_G$ is  a Marchenko-Pastur distribution. Since the Marchenko-Pastur distribution, by construction, corresponds to the first moment equal to 1, one does not need to renormalize the spectral density.   In the case of the second element of the product, i.e . the  operator $|U_1+U_2|$, its spectral properties come from the special  case of  the general formula  proven in~\cite{HAAGLAR},
\begin{eqnarray}
R_{|U_1+U_2+...+U_k |}(z)=k \frac{\sqrt{1+4z^2}-1}{2z} .
\label{sumU}
\end{eqnarray}
Therefore, using the properties of  the R-transform and choosing $k=2$, one gets $G_{|U_1+U_2|}(z)= \frac{1}{\sqrt{z^2-4}}$. Since we need the Green's function for the square of the modulus of $|U_1+U_2|$, we use the symmetry relation $G_{H^2}(z)=G_H(\sqrt{z})/\sqrt{z}$, which yields $G_{|U_1+U_2|^2}(z)= \frac{1}{\sqrt{z(z-4)}}$. 
Expanding the last relation for $z \rightarrow \infty$, we see, that the first moment is equal to 2. So final rescaling according to Eq.~(\ref{resc}) recovers the Green's function for the arcsine distribution
\begin{eqnarray}
G_{AS}(z)=\frac{1}{\sqrt{z(z-2)}} ,
\end{eqnarray} 
which completes the proof. The subscript AS stands for "arcsine", see Eq.~(\ref{arcsin}) below. 

The above construction is easily generalizable for the case of arbitrary long strings of powers of Ginibre ensembles and sums of unitary ensembles.
In such a case, one has to use Eq.~(\ref{sumU}) and the fact, that in the large $N$ limit, the limiting spectral density of the product of $m$ identically distributed isotropic unitary random matrices is equal to  the spectral density of the $m$-th power of a  single matrix from such ensemble~\cite{BURNOWSW,AGT10}.  This observation allows  to recover spectral properties of  the generalizations of the Bures measures proposed in Refs.~\cite{CNZ10,ZPNC11,Ja12}. Further generalizations include the case of strings of  Marchenko-Pastur distributions for rectangular matrices $X$ and/or cases of truncated unitary distributions.  The general method is always based on the  Haagerup-Larsen theorem, but the final formulae are usually more involved compared to the case presented here. 
\section{Generalized Wishart matrices and their spectral densities}
\label{secIII}

\subsection{Quadratic equation}

As a warm-up exercise, we start by recalling simple problems
which  correspond to a quadratic equation.
Consider first the Green's function corresponding to the free binomial distribution, where
$\rho(\lambda)=\frac{1}{2}(\delta(\lambda)+\delta(\lambda-1))$. 
The Green's function reads therefore
$G(z)=\frac{1}{2}(\frac{1}{z}+\frac{1}{z-1})$.  
Straightforward manipulations yield the R-transform and S-transform,
 given respectively by $R(z)=(z-1+\sqrt{z^2+1})/(2z)$ and $S(z)=2(1+y)/(1+2y)$.
 Antici\-pating the results needed for the remaining  part of this work, 
we consider  now
 the free sum of two binomial distributions. Since the R-transform is additive, 
we get $R_{AS}(z)=2R(z)=(z-1+\sqrt{z^2+1})/z$. Then the corresponding S-transform reads
 $S_{AS}(z)=(z+2)/(2+2z)$.
Substituting it into Eq. (\ref{wz}) we get
\begin{equation}
 w z(w+2)= 2(1+w)^2 .
\label{arcs}
\end{equation}
Solving this for  $w$
we obtain two conjugated solutions. Selecting the
one with negative imaginary part and plugging it into Eq.~(\ref{rho})
yields the {\sl arcsine distribution},
\begin{equation}
AS(x) =\frac{1}{\pi} \frac{1}{\sqrt{x(2-x)}}\; , \  \ 
x\in [0,2].
\label{arcsin}
\end{equation} 
This distribution gives us the level density of the suitably normalized
 sum of a random unitary matrix $U$ and its adjoint $U^{\dagger}$.
It describes the ensemble of quantum states obtained by reduction
of a coherent combination of maximally entangled states \cite{ZPNC11}
and will be used here to construct other distributions.

Before moving on  to the cubic equation and more complicated cases,
let us recall how to obtain in this way the
 general form of the Marchenko--Pastur distribution.
It describes the asymptotic level density $\rho(x)$ of random states of 
 $\rho=XX^{\dagger}/{\rm Tr} XX^{\dagger}$,
where $X$ is a rectangular complex Ginibre matrix
of size $N \times M$. We choose the rectangularity parameter $c=M/N \leq 1$.  The case $c>1$ yields the same nonzero eigenvalues and,   additionally,  $N-M$ zero eigenvalues.  
Let us then start with the corresponding $S$--transform, $S_c(w)=1/(1+cw)$,
 which reduces to Eq.~(\ref{SMP}) for $c=1$.
Plugging this expression into (\ref{wz}) leads to a quadratic equation
\begin{equation}
zw = (1+w)(1+cw) .
\label{MPequ}
\end{equation}
Its solution with respect to $w$  with a negative imaginary part
together with Eq.~(\ref{rho}) allows one to obtain 
\begin{equation}
P_{1,c}(x)   =
       \frac{1}{2 \pi x c} \sqrt{(x-x_-)(x_+ - x)} ,
\label{MPfull}
\end{equation}
where  $x\in [x_-,x_+]$,
with the edges of the support at $x_{\pm}=1+c \pm 2\sqrt{c}$.
In the case $c \rightarrow 0$, the Marchenko-Pastur distribution reduces to $\rho(\lambda)=\delta(\lambda-1)$.


\subsection{Cubic equation and Cardano solutions}

Next, we are going to present  solutions of problems motivated 
by ensembles of random matrices,
for which Eq.~(\ref{wz}) becomes a cubic polynomial in $w=w(z)$.

\subsubsection{Fuss--Catalan distribution of order two}

To show the presented method in action, we
rederive the Fuss-Catalan distribution $P_2(x)=[P_1(x)]^{\boxtimes 2}$,
which describes ensemble Eq.~(\ref{wishart})
with $X$ being a product of two independent square Ginibre matrices.
As a starting point we thus take the square of the $S$ transform of the MP distribution,
$S_{FC_2}(w)= [S_{MP}(w)]^2=(1+w)^{-2}$.
Putting this form into Eq.~(\ref{wz}) we get a cubic equation
\begin{equation}
  wz=(1+w)^3 .
\label{eqfc2}
\end{equation}
Calculating the Green's   function Eq.~(\ref{Gz}) and making use of Eq.~(\ref{rho}) 
one obtains the Fuss--Catalan distribution of order two, 
\begin{equation}
P_2(x) =  \frac{\sqrt[3]{2} \sqrt{3}}{12 \pi} \;
 \frac{\bigl[\sqrt[3]{2} \left(27 + 3\sqrt{81-12x} \right)^{\frac{2}{3}} -
   6\sqrt[3]{x}\bigr] } {x^{\frac{2}{3}}
     \left(27 + 3\sqrt{81-12x} \right)^{\frac{1}{3}}},
\label{FC2}
\end{equation}
where $ x \in [0,27/4]$.
This result was first obtained in Ref.~\cite{PS01} in the context of 
the construction of generalized coherent states from combinatorial sequences,
and later used in Ref.~\cite{CNZ10} to describe the asymptotic level density 
of mixed quantum states related to certain graphs.

\subsubsection{Generalized Fuss--Catalan distribution $P_{2,c}$}
\begin{figure}[ht]
\centering
\scalebox{0.4}{\includegraphics{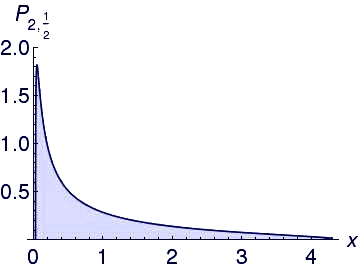}}
\caption{
Generalized Fuss-Catalan distribution of order two
$P_{2,c}(x)$
plotted for rectangularity parameter $c=1/2$.}
 \label{fig:fc2c}
\end{figure}

In  an analogous way we can treat the case of a product of two 
independent rectangular Ginibre matrices 
characterized by an rectangularity parameter $c=M/N$.
The corresponding $S$--transform $S_{2,c}=1/(1+cw)^2$
leads to a modified equation of the third order,  
\begin{equation}
  wz=(1+w)(1+cw)^2 .
\label{eqfc2c}
\end{equation}
Solving it with respect to $w$
and computing the corresponding Green's function Eq.~(\ref{Gz})
and its imaginary part one obtains a level density.
A particular case of the generalized Fuss-Catalan distribution
of order two obtained for $c=1/2$ is shown in Fig.~\ref{fig:fc2c}.
This precise  case was very recently studied in Ref.~\cite{DC14},
where an explicit density was provided.
Moments of the distributions $P_{s,c}$ can be expressed in terms
of the Fuss-Narayana numbers, see Ref.~\cite{MH2011}.

\subsubsection{Free--square root of the Marchenko--Pastur distribution}

To derive this distribution  we consider the square root of the $S$ transform of
the  MP distribution,  $S_{1/2}(w)= [S_{MP}(w)]^{1/2}$, which,  used Eq.~ (\ref{wz}),  yields a Cardano cubic equation,
\begin{equation}
w^3+(3-z^2)w^2+3w+1=0 .
\label{eqfcs2}
\end{equation}
Writing down the Green's function Eq.~(\ref{Gz}), we use Eq.~(\ref{rho}) 
to get an explicit form of the free multiplicative square
root of the Marchenko--Pastur distribution,  $P_{1/2}(x):=[P_1(x)]^{\boxtimes 1/2}$,
\begin{eqnarray}
P_{1/2}(x) & = 
 x^{-1/3}\frac{(9+Y)^{1/3} -(9-Y)^{1/3}}{2^{4/3} 3^{1/6} \pi} + \nonumber \\
& +  x^{1/3} \frac{(9+Y)^{2/3} -(9-Y)^{2/3}}{2^{4/3} 3^{5/6} \pi } ,
\label{eqfcs3}
\end{eqnarray}
where $Y(x)=\sqrt{81-12x^2}$
and  $x$ belongs to $[0,\sqrt{27/4}]$.

This distribution was derived in \cite{MPZ12}
using the inverse Mellin transform and the Meijer $G$ functions.
For the moment, we are not aware of any method
to generate an ensemble of random matrices
characterized asymptotically by the above level density.

\subsubsection{Bures distribution}

The Bures distribution describes the asymptotic level density
of random mixed states distributed according to the measure
\cite{SZ04} induced
by the Bures metric \cite{Uh92}. As already mentioned in section~II, to generate random states 
with respect to this measure it is sufficient \cite{OSZ10} 
to take $X=({\mathbbm 1}+U)G$, where $U$ is a Haar random unitary matrix
and $G$ is a square random Ginibre matrix of the same size,
 and substitute it into Eq.~(\ref{wishart}).
Using the Haagerup-Larsen theorem, we  have demonstrated  in section II why  the Bures distribution
can be represented as the multiplicative free product
of the positive arcsine law and the Marchenko-Pastur law: \
$B_1 = AS \boxtimes MP$.
The free $S$-transform of $B_1$ reads 
\begin{equation}
S_{B_1}(w)=\frac{w+2}{2(w+1)^2}=\frac{w+2}{2w+2}\cdot\frac{1}{1+w}.
\label{bur1}
\end{equation}
Observe that the first factor 
is the $S$-transform of $AS$
while the second one, $1/(1+w)$, 
is the $S$-transform of $MP$. This is the aforementioned  law of free multiplication.
The $S$ transform Eq.~(\ref{bur1})
together with  Eq. (\ref{wz}) leads to an equation of order three,
$w z(w+2)=2(1+w)^3$, which can be explicitly solved with respect to 
the complex variable $w$. Making use of Eqs.~(\ref{Gz}) and 
(\ref{rho}) one arrives at the Bures density 
\begin{equation}
B_1(x)= 
C
\left[ 
\left(\frac{a}{x} +\sqrt{\left(\frac{a}{x}\right)^2 \! \! -1} \right)^{2/3}
\!\!\!\!\!   -  \!
\left(\frac{a}{x} -\sqrt{\left( \frac{a}{x}\right)^2 \!  \! -1} \right)^{2/3}
\right]
\label{densitB}
\end{equation}
where $C=1/4\pi \sqrt{3}$ and $a=3\sqrt{3}$.
This distribution, first obtained in \cite{SZ04},
is defined on a support larger than the standard MP distribution, 
$x \in [0, a]$,  and it diverges for $x\to 0$ as $x^{-2/3}$.

\subsubsection{Generalized Bures distributions}

The generalized  Bures distribution  can be defined by a 
convolution of the arcsine and the Marchenko--Pastur
distribution with rectangularity parameter $c$,
namely  $B_{1,c}= AS \boxtimes P_{1,c}$. 
The corresponding ensemble
of random matrices can be obtained
writing $X=(1+U)G$ where 
$U$ stands for a random unitary matrix of size
$N$ generated according to the Haar measure
on $U(N)$, while $G$ denotes
a rectangular non-hermitian random Ginibre matrix
of order $N \times K$ with $c=K/N$.
Similar ensembles of random matrices were 
recently studied by Jarosz~\cite{Ja12}.
Getting  the corresponding ensemble of 
density matrices one may use
superpositions of pure states
of a four-party system followed by 
a projection on maximally entangled states
and taking a partial trace~\cite{ZPNC11}.

\begin{figure}[ht]
\centering
\scalebox{0.35}{\includegraphics{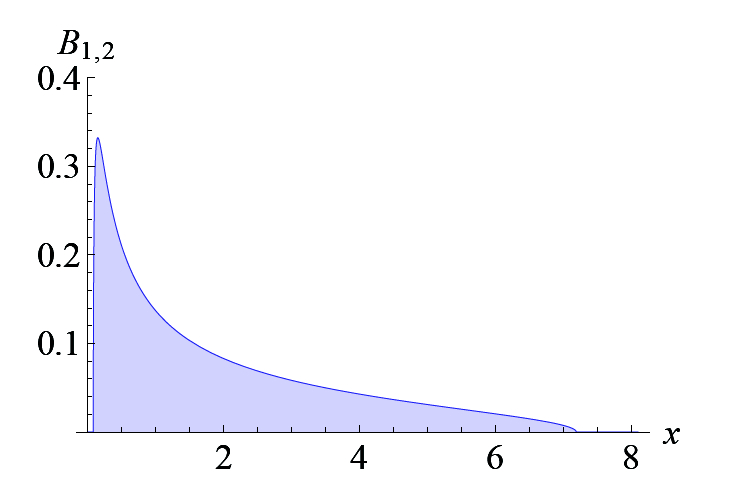}}
\caption{
The continuous part of the generalized Bures distribution $B_{1,c}(x)$
plotted for rectangularity parameter $c=2$,
so the shaded area equals $1/2$.}
 \label{fig:Bur2}
\end{figure}

\begin{figure}[ht]
\centering
\scalebox{0.35}{\includegraphics{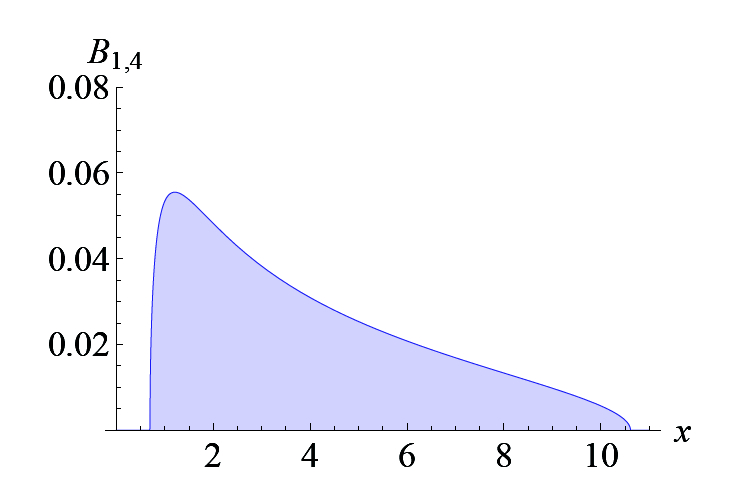}}
\caption{
As in Fig.~\ref{fig:Bur2} for $c=4$,
  so the area under the curve is $1/4$.}
 \label{fig:Bur4}
\end{figure}

Multiplying the corresponding $S$--transforms
we get 
$S_{B_{1,c}}(w)= (w+2)/(2(1+w)(1+cw))$
which leads to the following cubic equation:
$w z(w+2)= 2(1+cw)(1+w)^2$.
In the special case $c=1/2$, the above equation
simplifies to the quadratic one, $wz=(1+w)^2$,
corresponding to the Marchenko--Pastur distribution.
The generalized Bures distribution
$B_{1,c}(x)$ for $c\in [1/2,1]$
can be thus interpreted as an 
interpolation between the MP and Bures distributions.
In the case $c\le 1$ this distribution is absolutely 
continuous. In the case $c>1$, presented in
Figs. \ref{fig:Bur2} and \ref{fig:Bur4},
the distribution consists of a Dirac 
delta, $\delta(x)$ with weight $(1-1/c)$ 
and a continuous part -- see Th.~4.1 in~\cite{Be03}.

We shall conclude this section emphasizing that the method discussed
here is not limited to the cases presented. For instance,
analyzing the free multiplicative square root 
of the arcsine distribution, $AS^{\boxtimes 1/2}$,
or its free square, $AS^{\boxtimes 2}$,
one arrives at similar cubic equations, 
$(w+2)w^2z^2= 2(w+1)^3$, and
$(w+2)^2wz= 4(w+1)^3$, respectively,
which allow for the  derivation of  corresponding level densities.


\subsection{Quartic equation and Ferrari solutions}

The list of cases for which Eq.~(\ref{wz}) 
forms a quartic equation 
contains for instance,  the third order Fuss-Catalan distribution $P_3$,
the third root of the Marchenko Pastur distribution, $P_{1/3}$,
and the higher order Bures distribution.

\subsubsection{Fuss--Catalan distribution of order three}

\begin{figure}[ht]
\centering
\scalebox{0.48}{\includegraphics{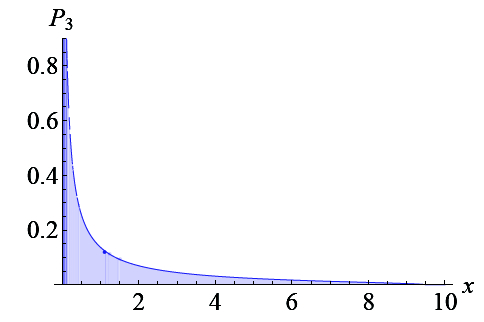}}
\caption{
Fuss-Catalan distribution $P_3(x)=[P_1(x)]^{\boxtimes 3}$
given in Eq. (\ref{eqfc3b}).}
 \label{fig:FC3}
\end{figure}

To find an analytical expression for the Fuss--Catalan distribution, 
$P_3=[P_1(x)]^{\boxtimes 3}$,
describing the asymptotic level density of the normalized Wishart matrix $XX^{\dagger}$,
where $X$ is a product of three independent Ginibre matrices,
we start with the third power of the $S$--transform corresponding to 
the Marchenko Pastur distribution, 
$S_3(w)=S_{MP}^3=1/(1+w)^3$.
Equation (\ref{wz}) leads then to the following  quartic equation:
\begin{equation}
w^4+4w^3+6w^2+w(4-z)+1=0 .
\label{eqfc3}
\end{equation}
Making use of the standard Ferrari formulae, 
we  obtain four explicit solutions of this
equation given as square roots of expressions
which contain polynomials of $z$ in powers $1/3$ and $-1/3$.
Analyzing the imaginary part of the corresponding Green's 
function Eq.~(\ref{Gz}), as discussed in  Appendix A,
we arrive at an explicit expression for the Fuss-Catalan distribution
of order three, 

\begin{equation}
P_3(x)= 
\frac{x^{-3/4}} {2 \cdot  3^{1/4} \pi }
\sqrt{ 4 Y  -  \frac{3^{3/4}  x^{1/4} } { \sqrt{Y} } }  ,
\label{eqfc3b}
\end{equation}
where $Y(x)=\cos \Bigl[ \frac{1}{3} \arccos \bigl( \frac{3 \sqrt{3}}{16} \sqrt{x} \bigr) \Bigr]$
and  $x  \in  [0,256/27]$.
Interestingly, the same distribution, shown in Fig.~\ref{fig:FC3} was derived 
earlier in  Ref.~\cite{PZ11}
and expressed in terms of combinations of hypergeometric functions $_3F_2(x)$,
which in this specific case admits a more  elementary representation.

Note that in an analogous way it is also possible 
to obtain expressions for the generalized
Fuss-Catalan distributions of order three, $P_{3,c}$,
which correspond to the $S$--transform
$S_{2,c}=1/(1+cw)^3$.
This distribution, representing the asymptotic level density of the Wishart
matrices obtained from a product of three independent
rectangular Ginibre matrices with rectangularity parameter $c=N/K$,
may in principle be further generalized for
three different rectangularity parameters, so that the
{\ $S$--transform} reads $S_{2,c}=1/(1+c_1w)(1+c_2w)(1+c_3w)$
-- see also \cite{DC14}.

\subsubsection{Free--third root of Marchenko--Pastur}

Consider the third root of the $S$ transform corresponding to 
the  MP distribution,  $S_{1/3}(w)=[S_{MP}(w)]^{1/3}$.
This choice applied to Eq.~(\ref{wz})
leads  again to a quartic equation in terms of $w$,


\begin{equation}
w^4+(4-z^3)w^3+6w^2+4w+1=0 .
\label{eqfsq3}
\end{equation}
Solving  this equation analytically for $w$, evaluating 
the Green's  function Eq.~(\ref{Gz}) and applying Eq.~(\ref{rho}) 
we arrive at the following form of the third free multiplicative 
root of the Marchenko--Pastur distribution,  $P_{1/3}(x):=[P_1(x)]^{\boxtimes 1/3}$,

\begin{equation}
P_{1/3}(x) = 
\frac{1}{2 \pi x}
\Bigl[ Y + 4x^3 -\frac{1}{2} x^6 +
\Bigl|  \frac{x^3 (24-12x^3+x^6)}{4 \sqrt{Y -2x^3+\frac{1}{4}x^6}} \Bigr|
\Bigr]^{1/2} ,
\label{eqfsq3b}
\end{equation}
where $Y(x)=(4/\sqrt{3}) x^{3/2}\cos \Bigl[ \frac{1}{3} \arccos \bigl( \frac{3 \sqrt{3}}{16} x^{3/2} \bigr)
 \Bigr]$
and  $x \in [0,  (256/27)^{1/3} ]$, see Fig.~4.

\subsubsection{2-Bures distributions}

The higher order $s$-Bures distribution
can be defined as a free convolution
of the arcsine and the $s$--Fuss--Catalan distribution,
$B_s=AS\boxtimes P_s$.
It describes the asymptotic
level density of the Wishart matrices $XX^{\dagger}$,
where $X=({\mathbbm 1}+U)G_1 \cdots G_s$. Here
$U$ denotes a random unitary matrix distributed according to the 
Haar measure while $G_1,\dots G_s$ are independent square 
complex Ginibre matrices. In the case $s=1$
 the standard Bures ensemble \cite{OSZ10} is retrieved.
Note that these distributions 
coincide with $\mu((s+2)/2,1/2)$ from Ref.~\cite{Ml10}, up to a dilation by 2.
Indeed, the free $S$-transform of $s$-Bures is
\begin{equation}
S(w)=\frac{w+2}{2(w+1)^{s+1}},
\end{equation}
which can be compared with (4.11)
in Ref.~\cite{Ml10} for $p=(s+2)/2$ and $r=1/2$.

Consider the case $s=2$ for which the Cauchy function
$S_{B_2}(w)= (w+2)/(2(1+w)^3)$
leads to the quartic equation
\[
wz (w+2) = 2(1+w)^4 .
\]
Next, out of four analytical Ferrari solutions
select  $w(z)=-1+(z+i\sqrt{(z-8)z})^{1/2}/2$,
which can be rewritten as $w=-1+\sqrt{8z} \exp[i \arccos(\sqrt{z/8})]$.
Plugging this into Eq.~(\ref{Gz}) we get the Green's function, which, 
used in Eq.~(\ref{rho}) yields the desired density,
$B_{2,1}(x)=\sin [\frac{1}{2} \arccos (\sqrt{2x}/4)]/(2^{1/4} \pi x^{3/4})$.
Making use of the known formula for  the sine of the half angle,
$\sin(x/2)=\sqrt{(1-\cos(x))/2}$ we can get rid of arc cosine 
and arrive at the result
\begin{equation}
B_2(x) =\frac{1}{\pi\; 2^{5/4} x^{3/4}} \sqrt{ 2 - \sqrt{x/2}},
\label{bur2}
\end{equation}
for $x \in [0,8]$, see Refs.~\cite{MPZ12} and \cite{For14}.
It is worth noting that other recent representations
of the Fuss-Catalan, Raney and related distributions \cite{HM12,Neu13,For14,GNS14}, 
also contain sine functions, 
the argument of which is an inverse trigonometric function
of the rescaled argument.

In a similar way one obtains results for the  generalized $2$--Bures
distribution $B_{2,c}(x)$, corresponding to the product 
$X=({\mathbbm 1}+U)G_1 G_2$ with rectangular matrices $G_1$ and $G_2$.
For any rectangularity parameter $c=N/M$, the corresponding
quartic equation reads now $wz(w+2)=2(1+cw)(1+w)^3$
and can be solved analytically. The  corresponding level densities
are too lengthy to be reproduced  here.
However, in the special case $c=1/2$
this equation reduces to the case of Eq.~(\ref{eqfc2}),
so the generalized $2$--Bures distribution 
with rectangularity parameter $c=1/2$ coincides with the
Fuss--Catalan distribution cf.~Eq.~(\ref{FC2}), $B_{2,1/2}(x)=P_2(x)$.

The list of other interesting cases that 
 lead to quartic equations includes, for instance,
the free multiplicative convolution of the arc-sine and Bures,
$AS{\boxtimes}B=AS^{\boxtimes 2}{\boxtimes}MP$,
or the  free multiplicative square root of the Bures distribution,
 $B^{\boxtimes 1/2}=AS^{\boxtimes 1/2} \boxtimes MP^{\boxtimes 1/2}$.
The corresponding level densities
can be obtained by solving quartic equations
$(w+2)^2 wz= 4(w+1)^4$ and $(w+2)w^2 z^2= 2(w+1)^4$, respectively.

\section{Concluding Remarks}

Making use of the $S$--transform and the Cauchy (Green's) function
it is possible to write down an explicit form
for the  probability measures defined by the free multiplicative convolution
of the Marchenko-Pastur (MP) distribution $P_1$ and other probability measures with known $S$-transform.
 For instance, the multiplicative convolution of the arcsine distribution and $P_1$
raised in the free multiplicative sense to an integer power leads to an algebraic equation for the argument of the $S$--transform. 
We studied some relevant cases
for which this algebraic equation is of the third or fourth order, so based on the 
known Cardano and Ferrari solutions, one can analytically derive  an explicit form of the
required probability measures. This is the case, for instance, 
for the free multiplicative powers of the Marchenko-Pastur distribution,  $[P_1(x)]^{\boxtimes s}$,
with an exponent $s$ equal to $2$ and $3$ and also $1/2$ and $1/3$,
and for the convolution of $P_1$ and $P_2(x)=[P_1(x)]^{\boxtimes 2}$
with the arcsine distribution ($AS$).
Among the  results of this work, it is worth mentioning the  
new analytic expressions for the densities Eqs.~(\ref{eqfcs3}), (\ref{eqfc3b}) and (\ref{eqfsq3b}).

Several distributions derived in this paper are useful in
the theory of random matrices and its numerous applications in physics.
Integer multiplicative powers of the MP, called Fuss--Catalan distributions,
describe the asymptotic level density of generalized Wishart random matrices, 
$W=XX^{\dagger}$, where  $X$ represents a product of $s$ independent
nonhermitian random square Ginibre matrices, $X=X_1\cdots X_s$.
We obtained here an explicit expression for $P_3=[P_1(x)]^{\boxtimes 3}$
in terms of elementary functions. We also  analyzed  the extension of the problem
for the case of rectangular Ginibre matrices. Furthermore,  
the case of the multiplicative convolution of AS with $P_1$ and $P_2$
corresponds to the Bures distribution $B_1$, 
the generalized Bures distribution $B_{1,c}$,  and the higher order Bures distribution $B_2$.
All of these  describe level distributions of generalized Wishart matrices $XX^{\dagger}$, where  $X$
is obtained  by multiplying the  sum of two random Haar  matrices with  a product of 
$s$ random Ginibre matrices. These results are applicable in the  description of the  asymptotic
level density of certain ensembles of random quantum states~\cite{ZPNC11}.

As a by-product of our analysis we derived explicit results for the probability measure
corresponding to the free multiplicative square/cubic  root of the Marchenko-Pastur distribution, written
$P_{1/2}=[MP]^{\boxtimes 1/2}$ and  $P_{1/3}=[MP]^{\boxtimes 1/3}$, respectively.
Note that for $p<1$ the distribution $[MP]^{\boxtimes p}$ is not
infinitely divisible with respect to the additive free convolution
$\boxplus$,
so the method of Cabanal--Duvillard \cite{CD05} is not applicable.
In fact, a stronger statement is true: if $p<1$ then the additive free power
$\left([MP]^{\boxtimes p}\right)^{\boxplus t}$
exists if  and only if $t\ge1$, see the recent result of 
Arizmendi and Hasebe \cite{AH14}.
It is thus unlikely  that there exists
a random matrix model which corresponds to the
level density described, for instance,  by the multiplicative free
square root of the Marchenko--Pastur distribution. 

\bigskip
Note added. 
After completing the paper, we became aware of two recent works where 
related issues of Raney-type distributions have been addressed using either differential 
equations~\cite{PEGO} or combinatorial analysis~\cite{For14}. 

\bigskip
Acknowledgements.
It is a pleasure to thank M.~Bo{\.z}ejko for his inspiring remarks,
encouragement and for inviting all of us for the Workshop
B{\c e}dlewo 2012, where this work was initiated.
We are also grateful to Z.~Burda and {\L}.~Rudnicki for fruitful interactions, 
and to P.~Forrester for communicating to us his recent paper~\cite{For14}
and several useful remarks on our paper. We also thank Martin Bier for the critical reading of the manuscript and 
numerous suggestions   leading to the improvement of the presentation. 
This work is supported  by the Grants DEC-2011/02/A/ST1/00119  (MN, K{\.Z}) and 
No. 2012/05/B/ST1/00626 (WM) of the Polish National Science Centre (NCN)
and in part by  the Transregio-12 project C4 of the Deutsche Forschungsgemeinschaft.
KAP acknowledges support from the PHC Polonium, France, project no. 28837QA.

\bigskip

\appendix
\section{On imaginary part of solution of a quartic equation}

To further demonstrate the derivation of the 
spectral density we treat in this appendix 
an example corresponding to the Fuss--Catalan distribution
of order three, cf. Eq.~ (\ref{eqfc3b}).
Writing down the Ferrari solutions of the quartic equation
(\ref{eqfc3}),  we identify the one with an negative 
imaginary part, denoted by $w_3$,
 so that the imaginary
part of the corresponding Green's  function Eq.~(\ref{Gz}) 
yields the desired spectral density Eq.~(\ref{rho}).

The full expression for this solution consists of two terms, $w_3(z)=a_1+a_2$.
We may omit the real
term $a_1$, as it does not contribute to the imaginary part of the Green's function.
The relevant term  then reads 
$$ a_2= -\frac{6^{2/3}}{12} 
\sqrt{- A-B + \frac{12 z}{\sqrt{A+B}}} $$,
where $z$--dependent symbols 
$A= \bigl(\frac{8z}{3z^2 +T /\sqrt{3}} \bigr)^{1/3}$
and  $B= (18z^2 + \sqrt{12} T)^{1/3}$
contain a square root $T=\sqrt{z^3(-256+27z)}$.
Its argument is negative for $ z \in [0,256/27]$,
so $T$  can be rewritten as $T= i \sqrt{z^3 (256-27z)}=it$,
where $t$ is a real number.
Let us now write the argument of the cubic root in $B$ 
in polar form, $Z=r e^{i \phi}$,
with radius $r=32\sqrt{3} z^{3/2}$ and phase $\phi=\arccos (3\sqrt{3} \sqrt{z}/16)$.
Then the key term reads
$$ a_2= -\frac{6^{2/3}}{12} 
\sqrt{- \frac{8z }{(Z/6)^{1/3}} - Z^{1/3}  + 
 \frac{12 z}{\sqrt{ \frac{8z}{(Z/6)^{1/3}} +Z^{1/3} } } }. $$
We can take the third root of $Z$ represented in polar form,
$Z^{1/3} =r^{1/3} \exp(i \phi/3)$,
group the terms $y \exp(i \phi/3)$  and  $ y \exp(- i \phi/3)$
and replace them by $2y \cos (\phi/3)$. Simplifying this 
 expression, we arrive eventually at the final 
form of the Green's  function Eq.~(\ref{Gz}) and
 by taking its imaginary part  Eq.~(\ref{rho}) 
we  arrive at the Fuss--Catalan distribution of order three 
Eq.~(\ref{eqfc3b}), defined for   $x\in [0,256/27]$.


\end{document}